# Free Electrons Can Induce Entanglement Between Photons

**Gefen Baranes, Ron Ruimy, Alexey Gorlach, and Ido Kaminer**

*Technion – Israel Institute of Technology, Haifa 32000, Israel*

kaminer@technion.ac.il

**Entanglement of photons is a fundamental feature of quantum mechanics, which stands at the core of quantum technologies such as photonic quantum computing, communication, and sensing. An ongoing challenge in all these is finding an efficient and controllable mechanism to entangle photons. Recent experimental developments in electron microscopy enable to control the quantum interaction between free electrons and light. Here, we show that free electrons can create entanglement and bunching of light. Free electrons can control the second-order coherence of initially independent photonic states, even in spatially separated cavities that cannot directly interact. Free electrons thus provide a type of optical nonlinearity that acts in a nonlocal manner, offering a new way of heralding the creation of entanglement. Intriguingly, pre-shaping the electron's wavefunction provides the knob for tuning the photonic quantum correlations. The concept can be generalized to entangle not only photons but also photonic quasiparticles such as plasmon-polaritons and phonons.**

## I. Introduction

Developed at the beginning of the twentieth century, quantum mechanics had by now changed the way we see the world. At the second half of the century, it gave birth to the fields of quantum optics and quantum information. One of the fundamental elements of quantum mechanics is the idea of entanglement, which is a nonlocal quantum correlation between different particles. This fundamental feature gave rise to numerous ideas in quantum technologies such as quantum teleportation [1], super dense coding [2], quantum cryptography [3–5], and quantum computation [6–9]. Photons are the fastest information carriers, have a very weak coupling to the environment, and thus are central to many areas of quantum technologies. Entanglement between two photonic states lay at the core of photonics-based quantum technologies, such as quantum sensing [10], imagining [11], metrology [12], cryptology [13], and photon-based quantum computers [14, 15]. Therefore, new ways to induce entanglement between photons or create entangled light sources have numerous applications in modern quantum technologies.

The conventional approach for creating entangled light uses linear optical operations such as beam splitters, or optical nonlinearities such as spontaneous parametric down-conversion (SPDC). Linear optical operations [16-18] cannot effectively entangle multi photon states, and techniques based on nonlinearities [19-23] are limited to specific wavelengths and typically suffer from low efficiencies. Thus, novel methods to create multiple-photon entangled states with non-trivial statistics have practical and fundamental importance. Our work proposes a novel way to entangle and bunch photonic states using *free electrons*.

The interaction of coherent free electrons with light can be described using the theory of photon-induced nearfield electron microscopy (PINEM) [24-30]. This interaction provides the ability to coherently shape a free electron's wavefunction [31–33] using femtosecond-pulsed lasers, as was shown experimentally in ultrafast transmission electron microscopes (UTEMs) [34–36]. This phenomenon was initially observed and described only for classical coherent-state light. However, recent works brought the electron-light interaction into the field of quantum optics [37–39], predicting theoretically [40–42] and demonstrating experimentally [43] electron–light interactions with different quantum state of light.

The interaction of a quantum light state and a free electron exchanges quantum information between the light and the electron. This idea has opened novel possibilities for using electrons to create novel quantum states of light [41], and a proposal to induce quantum correlations between different electrons interacting with the same photonic cavity [38]. Moreover, electrons with wavefunctions that are coherently shaped by light can also create entanglement of electrons with optical excitations [44, 45] or directly between separated qubits [46, 47]. These ideas show how free-electron interactions open novel research directions in quantum optics and quantum information.

However, from a fundamental perspective, all free-electron-based light-matter interactions so far were proposed to create entanglement in matter (qubits [46], free electrons [38]) rather than in light. This has left free-electron approaches outside an entire area of quantum optics: the creation of entangled light. We propose the use of free electrons as a new approach to address this fundamental challenge – creating desired states of multiple-photon entangled light.

Here we show that free electrons can entangle different photonic states (Fig. 1), even when they are spatially separated as with two different photonic cavities. Specifically, we find that post-selection of specific electron energies after the interaction enables creating even stronger entanglement. Importantly, the electron measurement enables heralding the successful creation of entanglement, as well as the bunching and anti-bunching of light. We find the underlying condition to enable the exchange quantum information: the electron's coherent energy uncertainty must be smaller than the energy of an individual photon. Using this condition, we propose a way by which free electrons can generate second-order correlations in light, making spatially separated light states bunched or anti-bunched. We propose a practical scheme – similar to the Hanbury Brown and Twiss (HBT) experiment [48] – that can be implemented in current setups (e.g., in UTEMs). There, we theoretically show how initially independent (i.e., uncorrelated) light sources can be correlated or anti-correlated using free electrons.

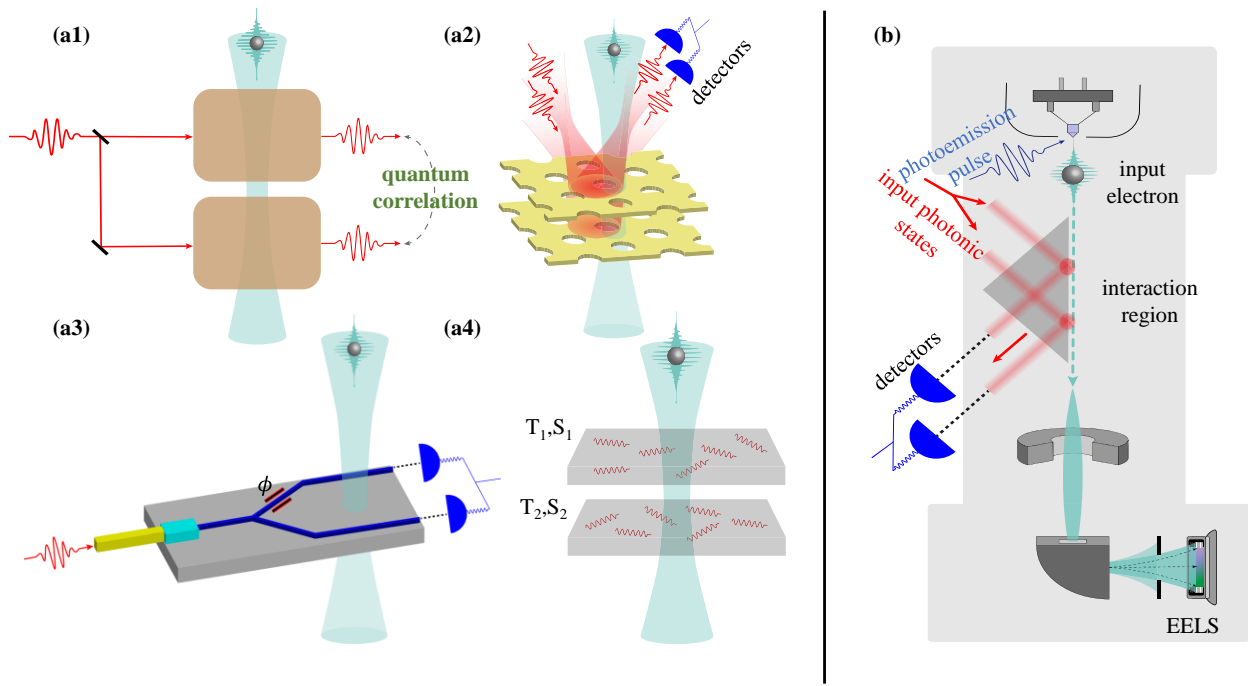

**Fig. 1. Free-electron interactions can create quantum correlations in different setups. (a)** The general concept (a1) can be implemented for electron interactions with different systems: photonic crystal cavities (a2), silicon-photonic waveguides (a3), and even other types of photonic quasiparticles [49] such as phonons (a4), which may entangle the temperatures of spatially separated objects. **(b)** One specific experimental setup that can demonstrate the general concept is the ultrafast transmission electron microscope (UTEM), where electrons can interact with evanescent fields and induce quantum correlations between these fields. Here, the two photonic states are created by having a light pulse split into two parts that undergo total internal reflection in two locations inside a prism (similar to [50]). Another relevant scheme was shown feasible in [29] using classical light.

Our approach has several attractive features connected with the nature of free electrons: electrons are ultrafast and thus enable creating quantum correlations between short pulses of light. Free electrons can also interact with any light frequency, are easy to detect [51] for post-selection and heralding of quantum light, and their wavefunction can be shaped [42, 49, 52] to provide additional degrees-of-freedom for controlling quantum correlations. In this manner, free electrons could be envisioned as *carriers of quantum information*, acting as a new type of *nonlinear* optical medium – one with a *nonlocal* optical response – with prospects to create phenomena such as quantum teleportation.

Looking at the bigger picture, our concept is directly applicable to the interaction of any quantum particle with any bosonic/photonic quasiparticle (e.g., plasmons, excitons, magnons, and phonons [49]). It is exciting to contemplate the broad range of systems that adhere to the general schemes we propose in Fig. 1; for example, electron-phonon interactions can be induced with the exact same formalism (Fig. 1a4), creating correlations between the entropy of phonons even in spatially separated solids, practically entangling their temperatures too.

## II. The theory of quantum interactions of free electrons with two photonic states

To model quantum electron–light interactions and their quantum correlations, we define our Hilbert space as the tensor product of $H_\text{e} \otimes H_\text{ph}$ and more generally $H_\text{e} \otimes H_{\text{ph}_1} \otimes H_{\text{ph}_2} \otimes ...$, with $H_\text{e}$ being the Hilbert space of the electron, and $H_{\text{ph}_i}, i = 1,2,...$ the Hilbert spaces of each photonic state. To account for the possibility of multiple electron interactions, we use the density-matrix formalism that includes mixed photonic states that cannot be described by a wavefucntion. In the case of two photonic cavities, we describe the initial state using the following density matrix:

$$\rho^{(i)} = \sum_{\substack{k,k'=-\infty \\ n_1,n_1'=0 \\ n_2,n_2'=0}}^{\infty} \rho_{k,n_1,n_2,k',n_1',n_2'} |k,n_1,n_2\rangle\langle k',n_1',n_2'| = \rho_e \otimes \rho_{ph_1} \otimes \rho_{ph_2}, \tag{1}$$

with the final equality based on the (realistic) assumption that the initial electron and photon states are separable before any interaction occurs. The basis of the photonic states consists of the Fock states $|n_1\rangle, |n_2\rangle$. The basis of the electron state consists of $|k\rangle$, being the state of the electron with energy $E_0 + \hbar\omega k$. The integer $k$ represents the displacement of the electron energy in units of the light energy quanta $\hbar\omega$, relative to the central energy $E_0$ (known as the zero-loss peak).

We assume that the electron's central energy is much larger than the photon energy $E_0 \gg \hbar\omega$ and that the electron energy uncertainty is smaller than the light quanta. These assumptions enable defining the electron states $|k\rangle$ as discrete. The assumptions are all satisfied in current electron microscopy experiments [29, 34, 43, 50] that have interactions with photonic excitations that are typically in the optical range. This is exactly the case in all theoretical and experimental papers in the field [24, 25, 30]. In the rest of the paper, we use this assumption to model all electrons as having discrete energies. This way, we can consider as initial conditions both conventional monoenergetic electron states $|0\rangle$ ("unshaped", the spectrum contains only a single energy peak), or "shaped" electron states that appear as coherent superpositions of multiple electron energy states.

The state of the system after one interaction with an electron is

$$\rho^{(f)} = S_2 U_\varphi S_1 \rho^{(m)} S_1^\dagger U_\varphi^\dagger S_2^\dagger, \tag{2}$$

where $U_\varphi$ is the free space propagation (FSP) operator which affects the electron state in the manner of $U_\varphi|E_k\rangle = e^{-i\varphi k^2}|E_k\rangle$. $\varphi$ is defined by $\varphi = 2\pi \frac{z}{z_D}$, with $z$ being the propagation distance and $z_D = 4\pi\gamma_0^3 m_e v^3/\hbar\omega^2$. $m_e$ is the electron mass, $v$ its velocity, and $\gamma_0$ its Lorentz factor. During FSP, each electron energy component accumulates phase at a different rate, as was demonstrated inside electron microscopes and other systems [29, 53, 54]. $S_1$ and $S_2$ are the scattering matrices describing the quantum electron–photon interaction with different photonic states [38]:

$$S_i = \exp[g_Q b a_i^\dagger - g_Q^* b^\dagger a_i], i = 1,2. \tag{3}$$

$a_i^\dagger, a_i$ are the photonic creation/annihilation operators; $b^\dagger, b$ describe the translation in momentum (or energy) of the electron, which can be defined as $b^\dagger|k\rangle = |k+1\rangle$, $b|k\rangle = |k-1\rangle$, or equivalently as $b = \exp(i\omega z/v)$ [49]; $g_Q$ is the coupling strength, which can be defined as an integral on the electric field along the electron trajectory (considering the second quantized field that corresponds to a single photon excitation) [28, 38].

Because the operators of the light and the electron commute, $[a, b] = 0$, we can represent the scattering matrices as a sum of the multiplication of two commuting operators – one for the light and one for the electron. Consequently, we develop an efficient representation of the interaction with multiple cavities:

$$S_2 U_\varphi S_1 = \sum_{k=-\infty}^{\infty} A_k^\varphi(a_1, a_2) b^k, \tag{4}$$

$$A_k^\varphi(a_1, a_2) = g_Q{}^k e^{|g_Q|^2} \sum_{k_2=-\infty}^{\infty} F_{k_2}(a_2) F_{k-k_2}(a_1) e^{-i\varphi(k-k_2)^2}, \tag{5}$$

$$F_k(a) = \sum_{m_2=0}^{\infty}\left[\frac{(-1)^{m_2}|g_Q|^{2m_2}a^{m_2}(a^\dagger)^{k+m_2}}{m_2!}\frac{1}{(k+m_2)!}\right]. \tag{6}$$

The $A_k$ operators contain only the photonic creation/annihilation operators $a_i^\dagger, a_i$. The complete derivation can be found in the Supplementary Material (SM) S.1.1.

This approach allows us to investigate and simulate many interactions of consecutive electrons with one or more photonic states. Next, we investigate induced correlations between two states of light for two different options: 1) when not measuring the electron's energy and tracing it out after each interaction; 2) when measuring the electron after the interaction and post-selecting on certain desired energy values.

### III. Inducing photon correlations using free electrons

In this section, we show how a free-electron interaction can correlate two states of light even without measuring the electron state after the interaction (i.e., without post-selection). A strong entanglement is possible when post-selecting on the electron energy measurement (shown in the next section). It was shown that electrons of narrow energy distribution (i.e., a single energetic state, a Kronecker "delta" electron, described by the zero-loss energy $|E_0\rangle$) can induce entanglement between the electron and a photonic state [38]. This discovery motivates utilizing electrons to create quantum correlations between the photonic cavities. In the language of quantum information, we regard the two photonic states as the *principal system* and the electron as the *environment* [55]. Since the electron does not have to be measured following the interaction, the system is generally an open quantum system. After each electron passed by, we trace it out to find the density matrix of the photonic system $\rho_{\text{ph}_1,\text{ph}_2}^{(m+1)} = \text{Tr}_e(\rho^{(m)}) = \sum_{k=-\infty}^{\infty}\langle k|\rho^{(m)}|k\rangle$. We then repeat the calculation with the next electron (an illustration is provided in Fig. 2a).

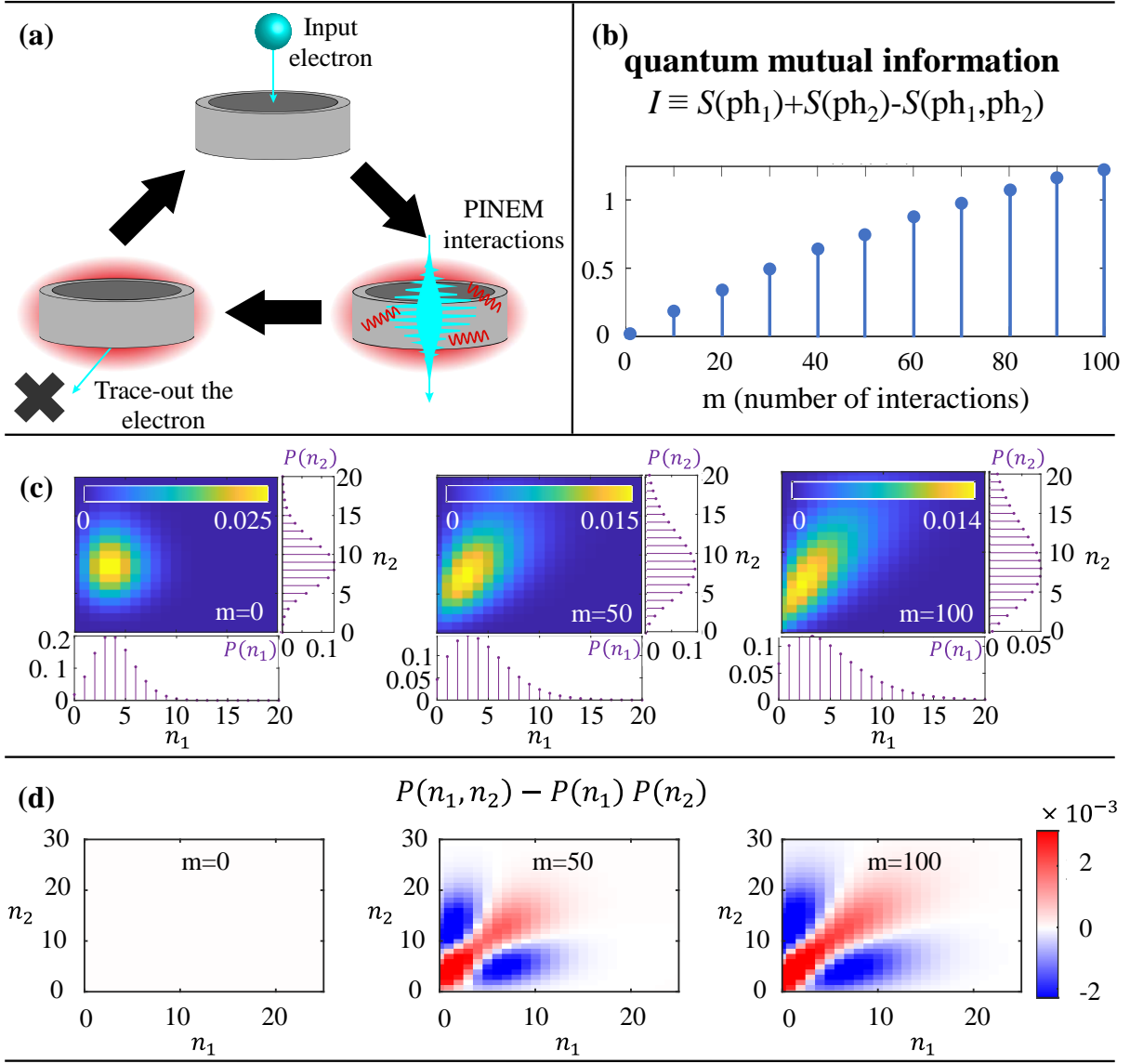

**Fig. 2. Correlations induced between two photonic states following one or more interactions with free electrons.** (a) Interaction scheme. After each interaction of an electron and the two photonic states, we trace out the electron. In certain cases, we then input another electron into the system and repeat. (b) The quantum mutual information as a function of the number of interactions increases with the number of interactions, indicating that correlations are induced. (c) Statistic of two coherent states with $\alpha_1 = 2$ and $\alpha_2 = 3$ after $m = 0, 50, 100$ consecutive interactions with free electrons. (d) $P(n_1, n_2) - P(n_1) \cdot P(n_2)$ after $m = 0, 50, 100$ consecutive interactions with free electrons increases and indicates correlations. The coupling strength in figures (b-d) is $g_Q = 0.1$.

The electron index that interacts with the light is represented by $m = 0,1,2,..$ with $m = 0$ being the initial state before any interaction (i.e., $|E_0\rangle$). The joint photonic state after $m$ consecutive electrons can be written as an operator-sum representation [55,56]:

$$\rho_{\text{ph}}^{(m+1)} = \text{Tr}_e\left[S_2 U_\varphi S_1 \rho_{\text{ph}}^{(m)} \otimes |\phi_e\rangle\langle\phi_e| S_1^\dagger U_\varphi^\dagger S_2^\dagger\right] = \sum_k \left(A_k(a_1, a_2) \rho_{\text{ph}}^{(m)} A_k^\dagger(a_1, a_2)\right), \quad (7)$$

while the $A_k$ operators, known as *operation* elements or *Kraus* operators, contain only the photonic creation/annihilation operators $a_i^\dagger, a_i$ and are defined in Eq. (5). The complete derivation can be found in SM S.1.2. Using this approach, one can describe the dynamics of the *principal system* (the two states of light) without explicitly considering the properties of the *environment* (the free electrons). This way we investigate the photonic states' properties, such as quantum correlations and statistics, after many interactions.

We examine the difference between the joint-photonic probability distribution $P(n_1, n_2)$ (which is depicted in Fig. 2c) and the multiplication of the individual photonic probability distributions $P(n_1)P(n_2)$. $P(n)$ is defined as the probability to have $n$ photons in the photonic state. The difference increases with each electron interaction, indicating correlations between the photonic states as depicted in Fig. 2d. We can also quantify the correlations using the *quantum mutual information* [55] between the two states of light (Fig. 2b), indicating that the photonic states become dependent on each other and increasingly so following more electron interactions. Note that these correlations are not necessarily quantum. Testing whether the correlations are truly quantum requires different criteria, as discussed below.

To check whether we have entanglement between the two photonic cavities in this case, we use two criteria from quantum information: the *Peres–Horodecki* criterion for entanglement [57, 58] and the *realignment* criterion [59, 60]. Both are sufficient but not necessary conditions for entanglement. Both criteria give a near-zero result, meaning that they could not prove or disprove the existence of entanglement. Further research on the amount of entanglement created by experiments such as we present here is a formidable task and left for future research. We show below that modifying the experiment to use electron post-selection avoids this inconclusive situation, creating states of light that are strongly entangled.

A key requirement for the transfer of information between the cavities is for the electron to undergo a certain distance of free space propagation (FSP) between the interactions. Without this evolution, electrons cannot mediate any influence of the first photonic state on the second one. We can explain this lack of influence by noticing that the absolute value of the electron's wavefunction is not modulated after a coherent electron-photon interaction [29] and rather requires FSP for the modulation to arise. Since the electron–light interaction only depends on the electron's amplitude (rather than its phase), the FSP stage is necessary. From a mathematical point of view, the key to the lack of induced correlations in the case of no FSP is that $S_1, S_2$ commute, $[S_1, S_2] = 0$ (SM S.2). Consequently, without FSP, there is complete symmetry between the two photonic states, and the theory does not change if one would replace the order of interactions or even if they both happen simultaneously.

To transfer quantum information between the cavities, we can shape the electron's amplitude in a manner that depends on the first cavity state, imprinting the information on the electron's wavefunction so it is transferred to the second cavity. Returning to the mathematical point of view, the addition of FSP breaks the symmetry in the problem, since the FSP operator does not commute with the $S_1, S_2$ operators.

The electron can give/take net energy to the second photonic state depending on the first photonic state, as shown in Fig. 3. It is interesting to consider whether such an effect can occur in the opposite direction, i.e., transferring energy backward through the induced entanglement. It is not trivial but possible to prove that regardless of the quantum state of the interacting electron, one cannot transfer energy backward. For the analytical proof and more information regarding this subject, see SM S.3. These results show how a natural *arrow of time* emerges from the *order of entanglement events*. Thanks to the specific commutation relations, the electron interaction can be described as occurring first with cavity $ph_1$ and afterward with cavity $ph_2$, creating a notion of causality as a result of the induced entanglement.

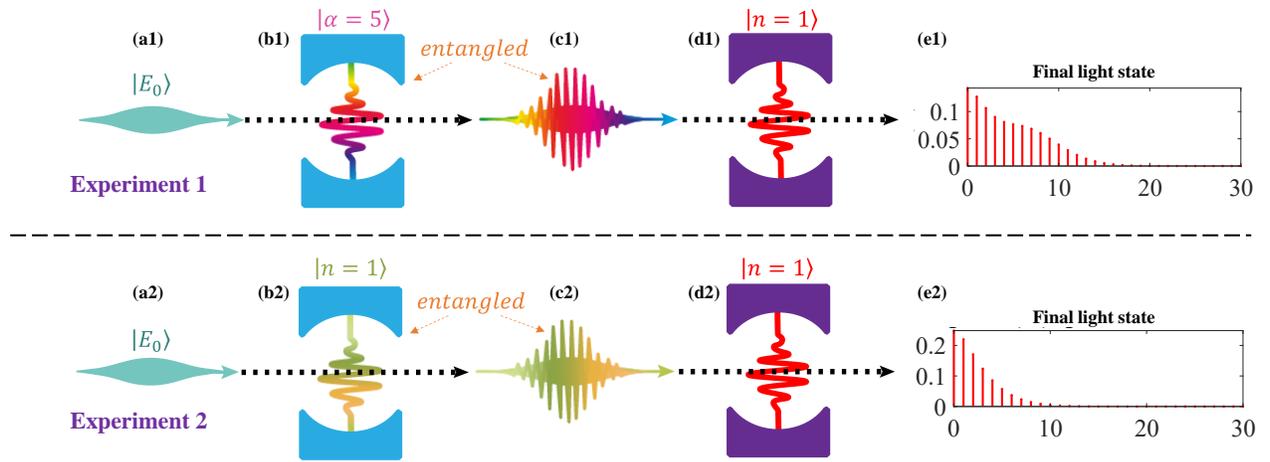

**Figure 3: The electron transfers quantum information between two photonic cavities through its entanglement with the cavities.** The schemes show two experiments with identical parameters, except for the first cavity's initial light state. The first interaction creates a joint entangled state of the electron and the cavity state. The second interaction transfers information to the second cavity. The first interaction can be understood as shaping of the electron wavefunction [42] that can then alter the statistics of the second cavity. **(a1,a2)** The same initial (unshaped) electron interacts with two cavities. **(b1,b2)** Each scheme presents a different light state in the first cavity. The initial light statistic in the first cavity: a coherent state with $\alpha = 5$ (b1) or a Fock state with one photon (b2). **(c1,c2)** The electron is shaped by the light state in the first cavity. **(d1,d2)** The light statistic in the second cavity for both experiments is a Fock state with one photon. It is modified in a way that depends on the state of light in the first cavity. **(e1,e2)** The final light statistics in the second cavity, plotted after 20 consecutive electron interactions, showing a clear difference that arises from the difference in light statistics in the first cavity. The interaction strength is $g_Q = 0.2$.

**IV. Inducing strong entanglement between two photonic states using post-selection**

Strong quantum correlations can be induced by post-selecting on the electron's energy after one interaction (as proposed recently for the generation of entanglement between qubits [46]). This approach allows the creation of quantum entanglement between many photons and even creating Bell states of light. This can be realized inside transmission electron microscopes because of the precision of the electron energy spectrometer that can measure a single event and distinguish the number of photons exchanged.

We begin with an electron at the zero-loss energy $|E_0\rangle$ that interacts with two arbitrary states of light as seen in Fig. 4a. After one interaction, we measure the electron with a projection operator of one defined energy, $E_0 - k\hbar\omega$, where $k$ is an integer. Using the $A_k$ operators defined above in Eq. (5), the density matrix of the remaining light state after the post-selection can be expressed as

$$\rho_{\text{ph}}^{(1)} = \frac{A_k(a_1, a_2)\rho_{\text{ph}}^{(0)} A_k^\dagger(a_1, a_2)}{P(k)}, \tag{8}$$

where $\rho_{\text{ph}}^{(0)} = \rho_{\text{ph1}}^{(0)} \otimes \rho_{\text{ph2}}^{(0)}$ is the initial density matrix of the two states of light. The probability to post-select the electron with the requested energy is (SM S.1.3)

$$P(k) = \text{Tr}\left(\rho_{\text{ph}}^{(0)} A_k^\dagger(a_1, a_2) A_k(a_1, a_2)\right) \tag{9}$$

After the measurement, the remaining joint light state is a pure state. In this case, we can quantify the entanglement between the two states of light by using the *entropy of entanglement* [61, 62]. The entropy of entanglement is the *von Neumann entropy* [63] of the reduced density matrix for any of the subsystems: For a pure state $\rho_{AB} = |\psi\rangle\langle\psi|_{AB}$, $S(\rho_A) = -\text{Tr}(\rho_A \log \rho_A) = S(\rho_B)$. In our case, the two subsystems are the two states of light. If the entropy of entanglement is non-zero, i.e., the subsystem is in a mixed state, it indicates that the two subsystems are entangled.

Fig. 4 shows that the free-electron interaction can induce quantum entanglement between any two general states of light, even ones with multiple photons. In addition, the electron is heralding the creation of entanglement and its amount when it is detected, providing a non-destructive measurement method for heralding of the entangled light. Specifically, when measuring the electron with electron energy loss spectroscopy (EELS), one can tell precisely which photonic parameters to expect and when the entanglement is occurring with femtosecond time resolution [24, 29].

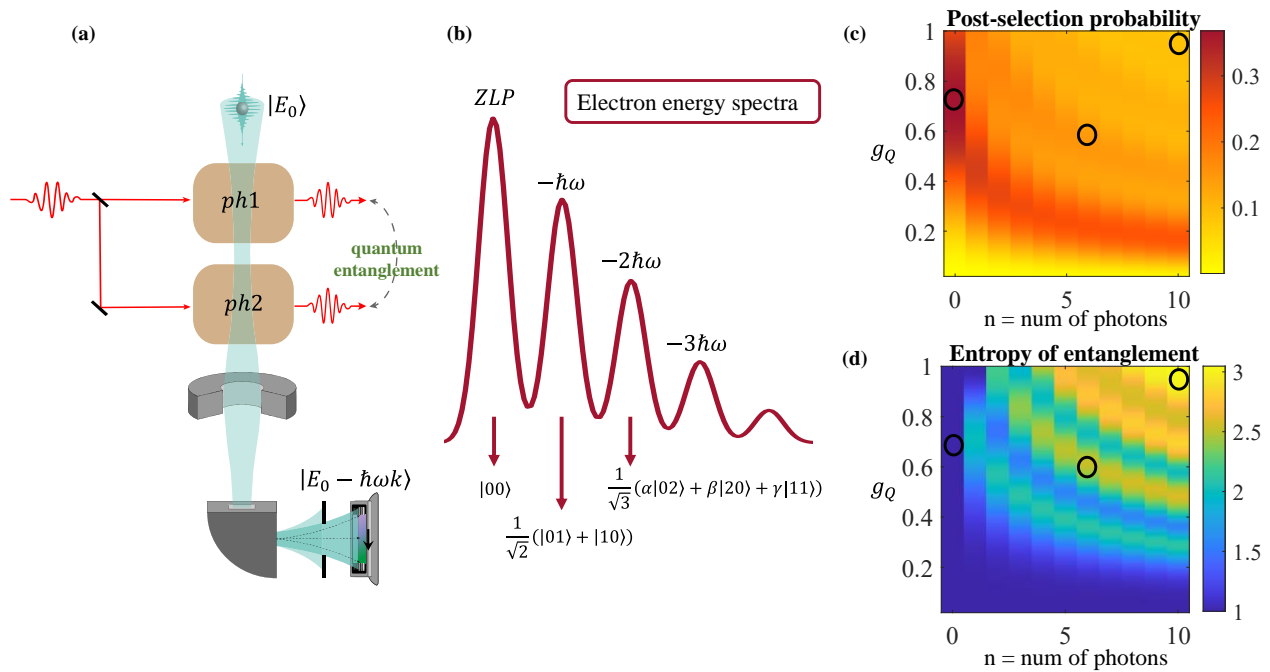

**Figure 4. Post-selection of electron energy to induce quantum entanglement between two photonic states.** (a) A scheme of the proposed experiment, where an (unshaped) electron interacts with two Fock states of light and is then measured with an electron energy spectrometer. The electron spectrometer enables post-selection of events in which an integer number of photons $k$ were emitted by the electron. (b) Each electron energy is entangled to a different photonic state after the interaction. (c) A map of the probability to post-select the electron with energy $E_0 - \hbar\omega$

(i.e., $k = 1$) after one interaction for different initial parameters. The x-axis is the number of photons in each photonic Fock state. The y-axis is the interaction strength $g_Q$. **(d)** A map of the *entropy of entanglement* between the two states of light after post-selecting on electrons with energy $E_0 - \hbar\omega$ (i.e., $k = 1$). The axes are the same as in (b). In both panels, we mark interesting initial parameters (one of them is discussed in the text and represents the creation of a Bell state).

Let us focus on a particular case of this result – the creation of a Bell state. An electron with initial energy $|E_0\rangle$ goes through two empty cavities and is later measured with energy $E_0 - \hbar\omega$. the post-selected state of light will be a Bell state $|\psi_{\text{ph}}\rangle = \frac{1}{\sqrt{2}}(|0\rangle_{\text{ph}_1}|1\rangle_{\text{ph}_2} + |1\rangle_{\text{ph}_1}|0\rangle_{\text{ph}_2})$. The probability of achieving this entangled state is $\sim 2|g_Q|^2$ for $|g_Q| \lesssim 0.2$, and the maximum value of this probability is $P = 0.37$, for $g_Q = 0.7$. this value of $g_Q$ can already be achieved in currents experiments [64]. Figs. 4c,d present maps of the post-selection probability and entanglement entropy as a function of $g_Q$ and the number of photons. Such experiments can be performed in current setups with repetition rates of few MHz [43], with higher repetition rates possible using different ultrafast microscopy techniques [65-67], and with continuous-wave operation recently shown possible [43, 68]).

## VI. Hanbury Brown and Twiss (HBT)-type experiment using free electrons

Until now, we have seen how one can use free electrons to create entanglement, with or without post-selecting them. Let us look at these experiments through the lens of quantum optics by imagining the following gedanken experiment: the initial photonic state is sent through a beam splitter into two interaction points with additional relative phase $\phi$. A second-order correlation between the two light states is detected, using the following definition of second-order correlations:

$$g^{(2)} = \frac{\langle n_1 n_2 \rangle}{\langle n_1 \rangle \langle n_2 \rangle}, \tag{10}$$

where $\langle \hat{a}_i \rangle = \text{Tr}(\rho_i a_i)$, $\langle n_i \rangle = \text{Tr}(\rho_i n_i)$, $\langle n_1 n_2 \rangle = \text{Tr}(\rho_{12} n_1 n_2)$ This measurement scheme is analogous to the Hanbury Brown and Twiss (HBT) experiment [48], except for the induced correlations arising from the free electrons' interactions. In addition, it gives a novel way to examine how free electrons can correlate or anticorrelate light.

We can consider the interaction of the same two states of light with multiple electrons if they arrive as a dilute beam so that we can neglect the coulombic interaction between them. Assuming the beam carries $m$ such electrons, and assuming that they are initially unshaped, the problem turns out to be analytical and straightforward using the Heisenberg picture. The initial second-order correlation function is $g_0^{(2)} = \frac{\langle n_1 n_2 \rangle^{(0)}}{\langle n_1 \rangle^{(0)} \langle n_2 \rangle^{(0)}}$. After $m$ electron interactions, we achieve the following second-order correlation (SM S.4):

$$g_N^{(2)} = \frac{g_0^{(2)} \langle n_1 \rangle^{(0)} \langle n_2 \rangle^{(0)} + N|g_Q|^2 \left( \langle n_1 \rangle^{(0)} + \langle n_2 \rangle^{(0)} + 2\text{Re}[\langle a_1^\dagger \rangle^{(0)} \langle a_2 \rangle^{(0)}] + (2N-1)|g_Q|^2 \right)}{\left( \langle n_1 \rangle^{(0)} + N|g_Q|^2 \right) \left( \langle n_2 \rangle^{(0)} + N|g_Q|^2 \right)}. \tag{11}$$

This formula shows that even for photonic states that were initially uncorrelated, i.e., $g_0^{(2)} = 1$, or for empty cavities, consecutive interactions can induce second-order correlations between them (Fig. 5a).

Interestingly, inserting FSP between the two interactions, we find that the FSP phase plays a similar role to the relative phase between the states of light (Fig. 5b3). Therefore, one can alternate between these two experimental knobs – creating the delay between the states of light by altering the actual light path or delaying the electron between the interactions (FSP). These ideas provide new degrees of freedom to shape the light statistics, induce second-order correlations, and create anti-bunched light.

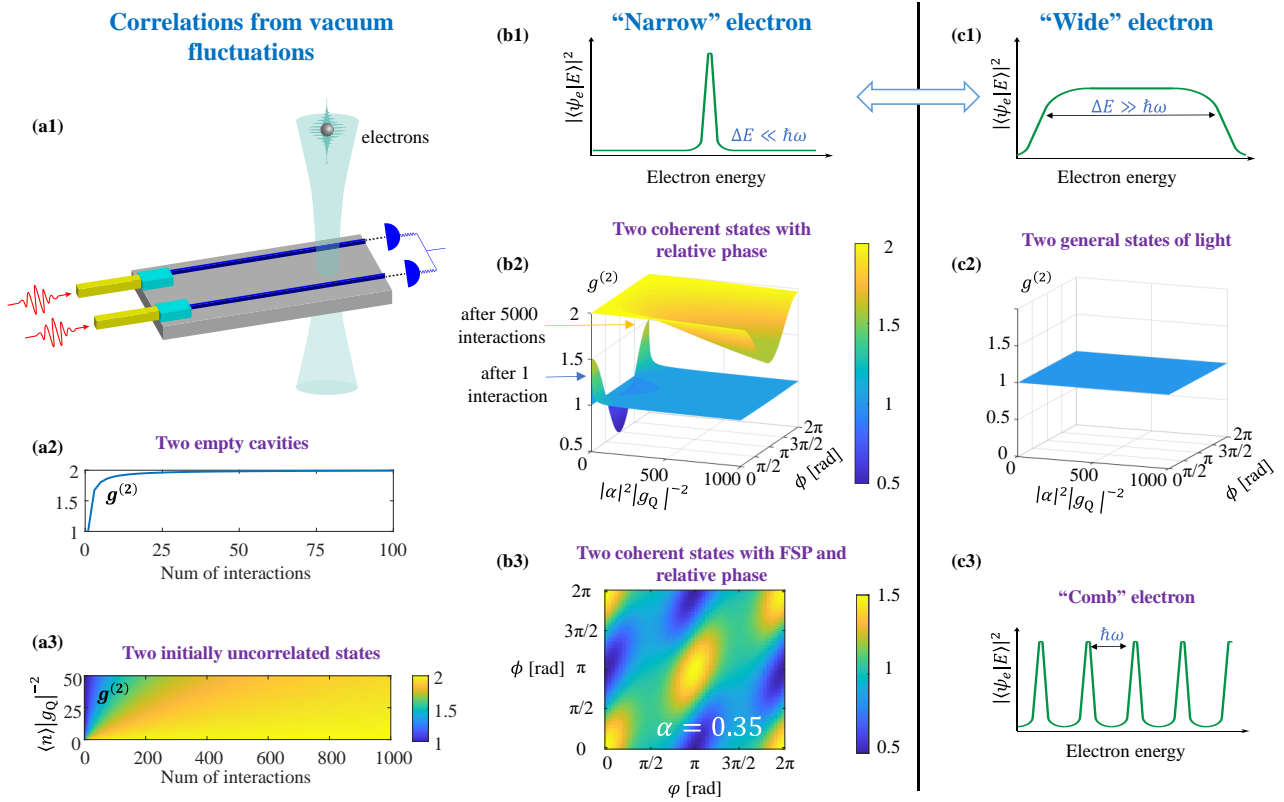

**Figure 5. Free electrons can bunch and anti-bunch light. The second-order coherence of light can be controlled by shaping the electron wavefunction.** An analogue of the Hanbury Brown and Twiss (HBT) experiment (a1), into which we add photon correlations through the interaction with free electrons. A scheme that can be implemented in current electron microscopes: two separated light pulses interact with consecutive (unshaped) electrons. $g^{(2)}$ for the light created in two initially empty cavities after electron interactions as a function of the number of electrons (a2), showing that second-order correlations can be induced when starting from vacuum fluctuations. Map of $g^{(2)}$ for any two initially uncorrelated states of light (a3). The result is independent of the specific states of light, only depending on the number of electrons and the average photon number (through the parameter $\langle n \rangle |g_Q|^{-2}$). For many electrons, $g^{(2)} \to 2$, the light becomes bunched, thermalized, and second-order correlations are induced. The comparison of electrons with narrow energy uncertainty (b1) and wide energy uncertainty (c1) shows that for a coherent uncertainty wider than the photon energy $\hbar\omega$, interaction cannot induce quantum correlations. The lack of correlations is seen in $g^{(2)}$ in panel (c2). It is possible to imitate the effect of a wide-uncertainty electron by instead creating a "comb" electron (c3) [38, 50]. The second-order correlation $g^{(2)}$ can be modified by consecutive interactions as shown in panel (b2), starting by two classical coherent states of light, and plotting $g^{(2)}$ after $N = 1$ and $N = 5000$ interactions. For large $N$, $g^{(2)} \to 2$, the light becomes bunched and thermalized. For a smaller number of interactions, it is possible to achieve $g^{(2)} < 1$, an anti-bunched light, depending on the relative phase $\phi$ between the two states of light. $\alpha$ is the coherent state parameter and $g_Q$ is the coupling strength. A map of $g^{(2)}$ after one interaction is shown in panel (b3), demonstrating the role of the relative phase of light $\phi$ and the phase $\varphi$ accumulated by the

free-space propagation (FSP) of the electron between the two points of interaction. The interaction strength is $g_Q = 0.4$.

**V. Pre-shaping the electron wavefunction to control correlations and to shape the light**

We found that the induced correlations strongly depend on the shape of the electron wavefunctions and the quantum state of the light. An electron cannot induce quantum correlations when its coherent energy uncertainty is comparable or bigger than the photon energy (compare Fig. 5b1 and 5c1). This concept is analogous to a similar uncertainty-width-dependence found in electron radiation mechanisms [69]. Electrons of wide coherent energy uncertainty, or equivalently, "comb" electrons that are made from a series of energy peaks coherent with one another (Fig. 5c3, as demonstrated recently [50]), cannot induce second-order correlations. Instead, their interaction with each cavity is equivalent to performing a displacement operator on the state of light in that cavity [38, 41], maintaining a separable joint state.

In contrast to the interactions of wide-uncertainty electrons, narrow-uncertainty electrons can be used to control the quantum state of light and achieve non-trivial photon statistics. When starting from an initially classical (coherent state) light, the electron can modify the correlations by creating anti-bunching (i.e., $g^{(2)} < 1$). For this to occur, we must set a certain delay that creates a relative phase difference $\phi$ between the two states of light, as shown in Fig. 5b2. Another way of achieving the relative phase needed for anti-bunching is delaying the electron between the interactions, as shown in Fig. 5b3.

**VI. Discussion and outlook**

Our work proposes a new way to create entanglement, bunching, and anti-bunching of light. We predict that coherent free electrons can entangle two spatially separated states of light, with the electron's energy uncertainty altering the level of entanglement. From a fundamental perspective, this shows that electron–photon interactions can create a novel type of nonlinearity, which is nonlocal and creates quantum correlations between separate photonic quasiparticles. Our proposed concepts can be implemented in state-of-the-art experiments such as in ultrafast transmission electron microscopy, which implement coherent electron–photon interactions [24-30, 36].

We emphasize that post-selecting on the electron's energy enables the creation of strong entanglement of photonic states, heralded by the measurement of the electron. Furthermore, pre-shaping of the electron's quantum state can control the quantum correlations. This way, one can use the electron to control the quantum shape of light and modify classical light sources into quantum. Our findings provide a step towards a bigger goal: inventing novel methods to create many-photon entangled states with non-trivial statistics. This approach is of fundamental interest for a wide range of applications of quantum technologies, such as photonic quantum computing.

We found that the electron can transfer information between two photonic states through its entanglement with both of them, in a process akin to quantum-teleportation. This process is possible only if the distance between the cavities is large enough such that electron dispersion has time to alter the electron's spatial shape. In that case, the order of interactions affects the result in a way that spawns a natural *arrow of time* in the problem. As an example of its consequences, when the distance between interactions is negligible (equivalent to a simultaneous interaction), the electron cannot exchange any information or energy between the cavities. The distance between interactions limits the transfer of information between the cavities.

These results can be extended beyond photonic cavities, for example, to electron–phonon interactions and to electron–qubit interactions, generalizing the results of previous works [46, 47]. Looking forward, we envision future quantum optical experiments based on networks of electrons and photons, creating high-dimensional hyper-entanglement that involve cluster states of spatially separated photonic states.

**References:**

[1] Nilsson, J. *et al*. Quantum teleportation using a light-emitting diode. *Nature Photonics*. **7**, 311–315 (2013).

[2] Harrow, A., Hayden, P., & Leung, D. Superdense coding of quantum states. *Physical Review Letters*. **92**, 187901 (2004).

[3] Ekert, A. K. Quantum Cryptography Based on Bell's Theorem. *Physical Review Letters*. **67**, 661 (1991).

[4] Brendel, J., Gisin, N., Tittel, W., & Zbinden, H. Pulsed Energy-Time Entangled Twin-Photon Source for Quantum Communication. *Physical Review Letters*. **82**, 2594 (1999).

[5] Beveratos, A., Brouri, R., Gacoin, T., Villing, A., Poizat, J. P., & Grangier, P. Single Photon Quantum Cryptography. *Physical Review Letters*. **89**, 187901 (2002).

[6] Menicucci, N. C., van Loock, P., Gu, M., Weedbrook, C., Ralph, T. C., & Nielsen, M. A. Universal quantum computation with continuous-variable cluster states. *Physical Review Letters*. **97**, 110501 (2006).

[7] Kok, P., Munro, W. J., Nemoto, K., Ralph, T. C., Dowling, J. P., & Milburn, G. J. (2007). Linear optical quantum computing with photonic qubits. *Reviews of Modern Physics*, **79**, 135 (2007).

[8] Aspuru-Guzik, A., & Walther, P. Photonic quantum simulators. *Nature Physics*. **8**. 285-291 (2012).

[9] Khasminskaya, S., Pyatkov, F., Słowik, K., Ferrari, S., Kahl, O., Kovalyuk, V., … Pernice, W. H. P. (2016). Fully integrated quantum photonic circuit with an electrically driven light source. *Nature Photonics*. **10**, 727–732 (2016).

[10] Degen, C. L., Reinhard, F., & Cappellaro, P. Quantum sensing. *Reviews of Modern Physics*. **89**, 035002 (2017).

[11] Lugiato, L. A., Gatti, A., & Brambilla, E. Quantum imaging. *J. Opt. B: Quantum Semiclass. Opt*. **4**, S176 (2002).

[12] Giovannetti, V., Lloyd, S., & MacCone, L. Advances in quantum metrology. *Nature Photonics*. **5**, 222-229 (2011).

[13] Gisin, N., Goire Ribordy, G., Tittel, W., & Zbinden, H. Quantum cryptography. *Rev. Mod. Phys*. **74**, 145 (2002).

[14] Braunstein, S. L., & van Loock, P. Quantum information with continuous variables. *Rev. Mod. Phys*. **77**, 513 (2005).

[15] Zhong, H.-S. *et al*. Quantum computational advantage using photons. *Science*. **370**, 6523 (2020).

[16] Li, X., Voss, P. L., Sharping, J. E., & Kumar, P. Optical-fiber source of polarization-entangled photons in the 1550 nm telecom band. *Physical Review Letters*. **94**, 053601 (2005).

[17] Krischek, R., Wieczorek, W., Ozawa, A., Kiesel, N., Michelberger, P., Udem, T., & Weinfurter, H. Ultraviolet enhancement cavity for ultrafast nonlinear optics and high-rate multiphoton entanglement experiments. *Nature Photonics*, **4**, 170–173 (2010).

[18] Zhang, J., Wildmann, J. S., Ding, F., *et al*. High yield and ultrafast sources of electrically triggered entangled-photon pairs based on strain-tunable quantum dots. *Nature Communications*. **6**, 10067 (2015).

[19] Kiesel, N., Schmid, C., Tóth, G., Solano, E., & Weinfurter, H. Experimental observation of four-photon entangled dicke state with high fidelity. *Physical Review Letters*. **98**, 063604 (2007).

[20] Valencia, A., Ceré, A., Shi, X., Molina-Terriza, G., & Torres, J. P. Shaping the waveform of entangled photons. *Physical Review Letters*. **99**, 243601 (2007).

[21] Baek, S. Y., Kwon, O., & Kim, Y. H. Temporal shaping of a heralded single-photon wave packet. *Physical Review A*. **77**, 013829 (2008).

[22] Mosley, P. J., Lundeen, J. S., Smith, B. J., Wasylczyk, P., U'Ren, A. B., Silberhorn, C., & Walmsley, I. A. Heralded generation of ultrafast single photons in pure quantum states. *Physical Review Letters*. **100**, 133601 (2008).

[23] Wieczorek, W., Schmid, C., Kiesel, N., Pohlner, R., Gühne, O., & Weinfurter, H. Experimental observation of an entire family of four-photon entangled states. *Physical Review Letters*. **101**, 010503 (2008).

[24] Barwick, B., Flannigan, D. J., & Zewail, A. H. Photon-induced near-field electron microscopy. *Nature*. **462**, 902–906 (2009).

[25] Garcia De Abajo, F. J., Asenjo-Garcia, A., & Kociak, M. Multiphoton absorption and emission by interaction of swift electrons with evanescent light fields. *Nano Letters*. **10**, 1859–1863 (2010).

[26] García De Abajo, F. J., Barwick, B., & Carbone, F. Electron diffraction by plasmon waves. *Physical Review B*, **94**, 041404 (2016).

[27] Piazza, L., Lummen, T. T. A., Quiñonez, E., Murooka, Y., Reed, B. W., Barwick, B., & Carbone, F. Simultaneous observation of the quantization and the interference pattern of a plasmonic near-field. *Nature Communications*. **6**, 6407 (2015).

[28] Feist, A., Echternkamp, K. E., Schauss, J., Yalunin, S. v., Schäfer, S., & Ropers, C. Quantum coherent optical phase modulation in an ultrafast transmission electron microscope. *Nature*. **521**, 7551 (2015).

[29] Priebe, K. E., Rathje, C., Yalunin, S. v., Hohage, T., Feist, A., Schäfer, S., & Ropers, C. Attosecond electron pulse trains and quantum state reconstruction in ultrafast transmission electron microscopy. *Nature Photonics*. **11**, 793–797 (2017).

[30] Park, S. T., Lin, M., & Zewail, A. H. Photon-induced near-field electron microscopy (PINEM): Theoretical and experimental. *New Journal of Physics*. **12**, 123028 (2010).

[31] Vanacore, G. M., Madan, I., Berruto, G., Wang, K., Pomarico, E., Lamb, R. J., … Carbone, F. Attosecond coherent control of free-electron wave functions using semi-infinite light fields. *Nature Communications*. **9**, 2694 (2018).

[32] Vanacore, G. M. *et al.* Ultrafast generation and control of an electron vortex beam via chiral plasmonic near fields. *Nature Materials*. **18**, 573–579 (2019).

[33] Echternkamp, K. E., Feist, A., Schäfer, S., & Ropers, C. Ramsey-type phase control of free-electron beams. *Nature Physics*, **12**, 1000–1004 (2016).

[34] Lobastov, V. A., Srinivasan, R., & Zewail, A. H. Four-dimensional ultrafast electron microscopy. *Proc Natl Acad Sci U S A*. **102**, 7069-73 (2005).

[35] Zewail, A. H. Four-Dimensional Electron Microscopy. *Science*. **328**, 5975 (2010).

[36] Polman, A., Kociak, M., & García de Abajo, F. J. Electron-beam spectroscopy for nanophotonics. *Nature Materials*. **18**, 1158-1171 (2019).

[37] Pan, Y., & Gover, A. Spontaneous and stimulated emissions of a preformed quantum free-electron wave function. *Physical Review A*. **99**, 052107 (2019).

[38] Kfir, O. Entanglements of Electrons and Cavity Photons in the Strong-Coupling Regime. *Physical Review Letters*. **123**, 103602 (2019).

[39] di Giulio, V., Kociak, M., & de Abajo, F. J. G. Probing quantum optical excitations with fast electrons. *Optica*. **6**, 1524-1534. (2019).

[40] Gorlach, A., Karnieli, A., Dahan, R., Cohen, E., Pe'er, A., & Kaminer, I. Ultrafast non-destructive measurement of the quantum state of light using free electrons. arXiv:2012.12069.

[41] Ben Hayun, A., Reinhardt, O., Nemirovsky, J., Karnieli, A., Rivera, N., & Kaminer, I. Shaping quantum photonic states using free electrons. *Sci. Adv*. **7**, eabe4270 (2021).

[42] Di Giulio, V., & García de Abajo, F. J. Free-electron shaping using quantum light. *Optica*. **7**, 1820 (2020).

[43] Dahan, R. *et al*. Imprinting the quantum statistics of photons on free electrons. arXiv:2105.03105 (2021).

[44] Mechel, C., Kurman, Y., Karnieli, A., Rivera, N., Arie, A., & Kaminer, I. Quantum correlations in electron microscopy. *Optica*. **8**, 70 (2021).

[45] Karnieli, A., Rivera, N., Arie, A., & Kaminer, I. Superradiance and Subradiance due to Quantum Interference of Entangled Free Electrons. *Physical Review Letters*. **127,** 060403 (2021).

[46] Zhao, Z., Sun, X.-Q., & Fan, S. Quantum entanglement and modulation enhancement of free-electron-bound-electron interaction. *Physical Review Letters.* **126**, 233402 (2020).

[47] Ruimy, R., Gorlach, A., Mechel, C., Rivera, N., & Kaminer, I. Towards atomic-resolution quantum measurements with coherently-shaped free electrons. *Physical Review Letters*. **126**, 233403 (2021).

[48] Hanbury Brown, R. & Twiss, R. Q. Correlation between photons in two coherent beams of light. *Journal of Astrophysics and Astronomy*. **15**, 13-19 (1994).

[49] Rivera, N., & Kaminer, I. Light–matter interactions with photonic quasiparticles. *Nature Reviews Physics*. **2**, 538-561 (2020).

[50] Dahan, R. *et al*. Resonant phase-matching between a light wave and a free-electron wavefunction. *Nature Physics.* **16**, 1123–1131 (2020).

[51] Egerton, R. F. Electron energy-loss spectroscopy in the TEM. *Reports on Progress in Physics*. **72**, 016502 (2008).

[52] Reinhardt, O., & Kaminer, I. Theory of Shaping Electron Wavepackets with Light. *ACS Photonics*. **7**, 2859–2870 (2020).

[53] Morimoto, Y., & Baum, P. Diffraction and microscopy with attosecond electron pulse trains. *Nature Physics.* **14**, 252–256 (2018).

[54] Kozák, M., Schönenberger, N., & Hommelhoff, P. Ponderomotive Generation and Detection of Attosecond Free-Electron Pulse Trains. *Physical Review Letters*. **120**, 103203 (2018).

[55] Nielsen, M. A., & Chuang, I. L. Quantum computation and quantum information. *Cambridge University Press.* (2010).

[56] Kraus, K., Böhm, A., Dollard, J. D., & Wootters, W. H. States, Effects, and Operations: Fundamental Notions of Quantum Theory. *Springer-Verlag*. (1983).

[57] Peres, A. Separability Criterion for Density Matrices. *Physical Review Letters*. **77**, 1413 (1996).

[58] Simon, R. Peres-Horodecki separability criterion for continuous variable systems. *Physical Review Letters*. **84**, 2726 (2000).

[59] Chen, K., & Wu, L.-A. A matrix realignment method for recognizing entanglement. *Quantum Inf. Comput.* **3**, 193-202 (2003).

[60] Nathaniel Johnston. QETLAB: A MATLAB toolbox for quantum entanglement, version 0.9. (2016).

[61] Vidal, G. Entanglement monotones. *Journal of Modern Optics*, **47**, 355–376 (2000).

[62] Vedral, V., Plenio, M. B., Rippin, M. A., & Knight, P. L. Quantifying Entanglement. *Physical Review Letters.* **78**, 2275 (1997).

[63] Petz, D. Entropy, von Neumann and the von Neumann entropy. *Springer-Dordrecht*. (2001).

[64] Adiv, Y., Hu, H., Tsesses, S., Dahan, R., Wang, K., Kurman, Y., Gorlach, A., Chen, H., Lin, X., Bartal, G. & Kaminer, I. 2D Cherenkov radiation and its photonic quantum nature. under review (2021).

[65] van Rens, J. F. M., Verhoeven, W., Kieft, E. R., Mutsaers, P. H. A., & Luiten, O. J. Dual mode microwave deflection cavities for ultrafast electron microscopy. *Applied Physics Letters*. **113**, 163104 (2018).

[66] Verhoeven, W., van Rens, J. F. M., Kieft, E. R., Mutsaers, P. H. A., & Luiten, O. J. High quality ultrafast transmission electron microscopy using resonant microwave cavities. *Ultramicroscopy*. **188**, 85–89 (2018).

[67] Verhoeven, W. *et al*. Design and characterization of dielectric filled TM110 microwave cavities for ultrafast electron microscopy. *Review of Scientific Instruments*. **90**, 083703 (2019).

[68] Henke, J.-W. *et al*. Integrated photonics enables continuous-beam electron phase modulation. arXiv:2105.03729 (2021).

[69] Karnieli, A., Rivera, N., Arie, A., & Kaminer, I. The coherence of light is fundamentally tied to the quantum coherence of the emitting particle. *Sci. Adv*. **7**, eabf8096 (2021).